\documentclass[3p,times,twocolumn]{elsarticle}

\usepackage{ecrc}

\volume{00}

\firstpage{1}

\journalname{Nuclear Physics B Proceedings Supplement}

\runauth{}

\jid{nuphbp}

\jnltitlelogo{Nuclear Physics B Proceedings Supplement}

\usepackage{amssymb,amsmath}

\usepackage[figuresright]{rotating}

\begin{document}

\begin{frontmatter}



\dochead{}

\title{Analyticity and higher twists}


\author[JINR]{Oleg Teryaev}
\ead{teryaev@theor.jinr.ru}

\address[JINR]{Bogoliubov Laboratory of Theoretical Physics, Joint Institute for Nuclear Research, 141980 Dubna, Russia}

\begin{abstract}
The representation for infinite sum of higher twists (HT) tower in
DIS implied by analyticity of virtual Compton amplitude is
suggested. Its simplest realization allows to describe the Bjorken
sum rule at all momentum transfers. It is stressed that TMDs
accommodate the infinite tower of HT similar to non-local vacuum
condensates for the case of vacuum matrix element. The D-term in
hadronic GPDs bears some similarity to vacuum cosmological constant.
The negative sign of D-term may be understood as a similarity
between inflation and annihilation via the graviton exchange.

\end{abstract}


\end{frontmatter}


\section{Introduction}

Higher twist (HT) corrections are very important for the applications of
QCD at low scales. The lower is the scale the higher twists enter
into the game. Sometimes any finite number of them is insufficient,
and account for infinite sums is necessary.

We address here several such cases. The first one corresponds to
(spin-dependent) DIS in real-photon limit, when contact with
low-energy theorems (GDH sum rules) may be achieved
\cite{Soffer:1992ck}. Originally, this was realized by matching of
the HT expansion in inverse powers of $Q^2$ (where QCD perturbative
expansion in $log Q^2$ was also included \cite{Soffer:2004ip}) and
"chiral" expansion in positive powers of $Q^2$. Here we describe
amother procedure, where HT series is represented in integral form,
incorporating the analytic properties of virtual Compton amplitude.
Even in its simplest version it leads to the rather accurate
description of Bjorken sum rule at all $Q^2$.

Another situation of all-twists relevance is represented by transverse momentum dependent parton distributions (TMDs).
They may be considered \cite{Ratcliffe:2007ye} as a (partial)sum of all HTs tower, while transverse moments (where Bessel moments \cite{Boer:2011xd} should naturally appear because transverse space is 2 -dimensional) correspond to definite twists.

The TMDs are therefore similar\cite{Teryaev:2004df} to the
non-local vacuum condensates\cite{Bakulev:1991ps} 
 also partially resumming the local
condesates of definite dimension playing the role of twist in the
vacuum. This may be an example of more general
similarity/universality of vacuum and hadronic matrix elements. It
is interesting, that vacuum analog of so-called Polyakov-Weiss
D-term \cite{Goeke:2001tz} in generalized parton distributions
(GPDs) is represented by nothing less than cosmological constant.
The observed definite (negative) sign of D-term may, in turn, be
interpreted as a positive "effective" cosmological constant in the
annihilation channel of respective gravitational formfactor
establishing the relation between inflation and annihilation.

\section{Resumming HT in spin-dependent DIS}
Let us consider as a case study the lowest non-singlet moment of
spin-dependent proton and neutron structure functions $g^{p,n}_1$
defined as
\begin{eqnarray}\label{eq1}
\Gamma_1^{p-n}(Q^2)=\int^1_0dx\, g^{p-n}_1(x,Q^2)\,,
\end{eqnarray}
with $x=Q^2/2M\nu$, the energy transfer $\nu$, and the nucleon mass
$M$. We imply the elastic contribution at $x=1$ to be excluded,
since the low-$Q^2$ behavior of ``inelastic'' $\Gamma^{p-n}_1(Q^2)$,
i.e. the Bjorken Sum Rule (BSR), is constrained by the
Gerasimov-Drell-Hearn (GDH) sum rule, which allows us to investigate
continuation of the Bjorken integral $\Gamma^{p-n}_1(Q^2)$ to low
$Q^2$ scales. As all the higher twists contributions are divergent
when $Q^2 \to 0$, only infinite sum may be matched to GDH value.
Let us consider the series
\begin{equation} S(Q^2)=\sum_1^\infty a_n ({M^2\over {Q^2}})^n.
\label{S}
\end{equation}
which may correspond either to non-perturbative part (HT) of
$\Gamma_1$ or to $I_1=2 M^2 \Gamma_1 /Q^2$ proportional to
photoabsorption cross-sections constrained by GDHSR. By
making the crucial step and representing $a_n$ as a moments
\begin{equation}
a_n = \int_{-\infty}^{\infty} f(x)x^{n-1}, \label{m}
\end{equation}
the sum of HTs can be recasted as
\begin{equation} S(Q^2)=
\int_{-\infty}^{\infty} dx {\frac{f(x) M^2}{Q^2 - x M^2}}.
\label{S1}
\end{equation}
Such a representation in terms of moments may be compared to the
similar one when the standard leading twist partonic expression is
derived. The later, besides the nice physical pictture, may be
justified by correct analytical properties of virtual Compton
amplitude, having s  and u cuts produced by respective poles (at LO)
and cuts in the partonic subprocess.

The similar arguments may be applied for HT resummation.
If analytical properties of $S(Q^2)$ are represented by the cut
residing at $Q^2 \leq 4 m^2$ (m being the mass of the lightest
particle in the respective channel) , the integration in (\ref{m})
should be limited to $(-\infty, - 4 m^2/M^2)$. If the function
$f(x)$ has a definite sign it leads to the alternating HT series.
Moreover, even for sign-changing $f(x)$ the series will be typically
alternating unless fine-tuning of $f$ is imposed.

It is crucially important that the leading twist contribution and
respective perturbative (logarithmic) corrections has the same
analytic properties, leading to  the same properties of the full
amplitude. This naturally selects the modified Analytic Perturbation
Theory (APT) \cite{Shirkov:1997wi} which was successfully applied
for the description of BSR  \cite{Pasechnik:2008th}. These dtudies
manifested the duality between the HT and perturbative corrrections,
so that HT decreased at NLO etc. Let me conjecture here, that the
"real" HT should correspond to the piece which cannot be absorbed to
perturbative series due to is asymptotic nature.

One may now continue $S$ to $Q^2=0$ which is defined by first
inverse moment
\begin{equation} S(0)=
- \int_{-\infty}^{-4 m^2/M^2} dx {\frac{f(x)}{x}}. \label{s0}
\end{equation}
Its sign will typically coincide with that of the first term of the
series.

The derivatives of $S$ at $Q^2=0$ are defined by higher inverse
momentss,say

\begin{equation}
S'(0)= - \int_{-\infty}^{-4 m^2/M^2} dx {\frac{f(x)}{M^2 x^2}}.
\label{s}
\end{equation}
If one neglect $m$ (which corresponds to minimal APT in the
perturbative part) this integral may diverge at $x \sim 0$. This
divergence may be used to cancel the infinite slope of minimal APT
contribution requiring that $f(x) \sim \rho_{pert}(s=M^2x) $ for $x
\sim 0$. This divergence is absent in the recently elaborated
Massive Perturbation Theory (MPT \cite{Shirkov:2012ux}, where $Q^2
\to \tilde Q^2 = Q^2+M^2_{gl}$) which together with the VDM form of
HT contribution $M_{HT}^2/(Q^2+M_{HT}^2)$ lead to the reasonable
description of BSR down to rather low $Q^2$.

Note that VDM form of HT perfectly fits to (\ref{S1}) with the
delta-function spectral density \footnote{It is interesting that
similar VDM form was discussed at this conference also in the talks
of Ya.Klopot for pion-photon transition formfactor \cite{Klopot:2013oua} (where it is
related to axial anomaly) and A. Aleksejevs for pion formfactor.}
while MPT expression has also the correct analytic properties
provided relevant "gluonic mass" $M_{gl} \geq \Lambda_{QCD}$.

At the same time, the attempt to match the MPT description with GDHSR fails. The reason is obvious: the (average)
slope of $\Gamma^{p-n}_1(Q^2)$  at low $Q^2$ is several times larger than the one following from GDHSR. This clearly supports the "two-component" approach \cite{Soffer:1992ck} where slope is decomposed to the sum  of "fast" rapidly decreasing
component due to structure function $g_2$ and "slow" component due to structure function $g_T=g_1+g_2$, which for BSR
provides the slope \cite{Soffer:2004ip} enhanced by factor $\mu^p_A/((\mu^n_A)^2-(\mu^p_A)^2) \sim 4$ determined by proton and neutron anomalous magnetic moments.

It is therefore natural to combine MPT analysis with approach
\cite{Soffer:1992ck}. The fast component contribution is controlled
by Butrkardt-Cottingham sum rule free from any corrections. GDH sum
rule allows to relate HT and gluon masses (appearing to be close) so
that there is single free parameter remained. The one-parameter
fits\footnote{I am grateful to I.Gabdrakhmanov and V. Khandramai for
the help in numerical calculations.} lead to the reasonable
description of the data with the quality increasing with taking into
account NLO MPT \cite{Shirkov:2012ux} and modifications of spectral
density (\ref{S1}),

\begin{figure}
\centerline{
\includegraphics[width=0.5\textwidth]{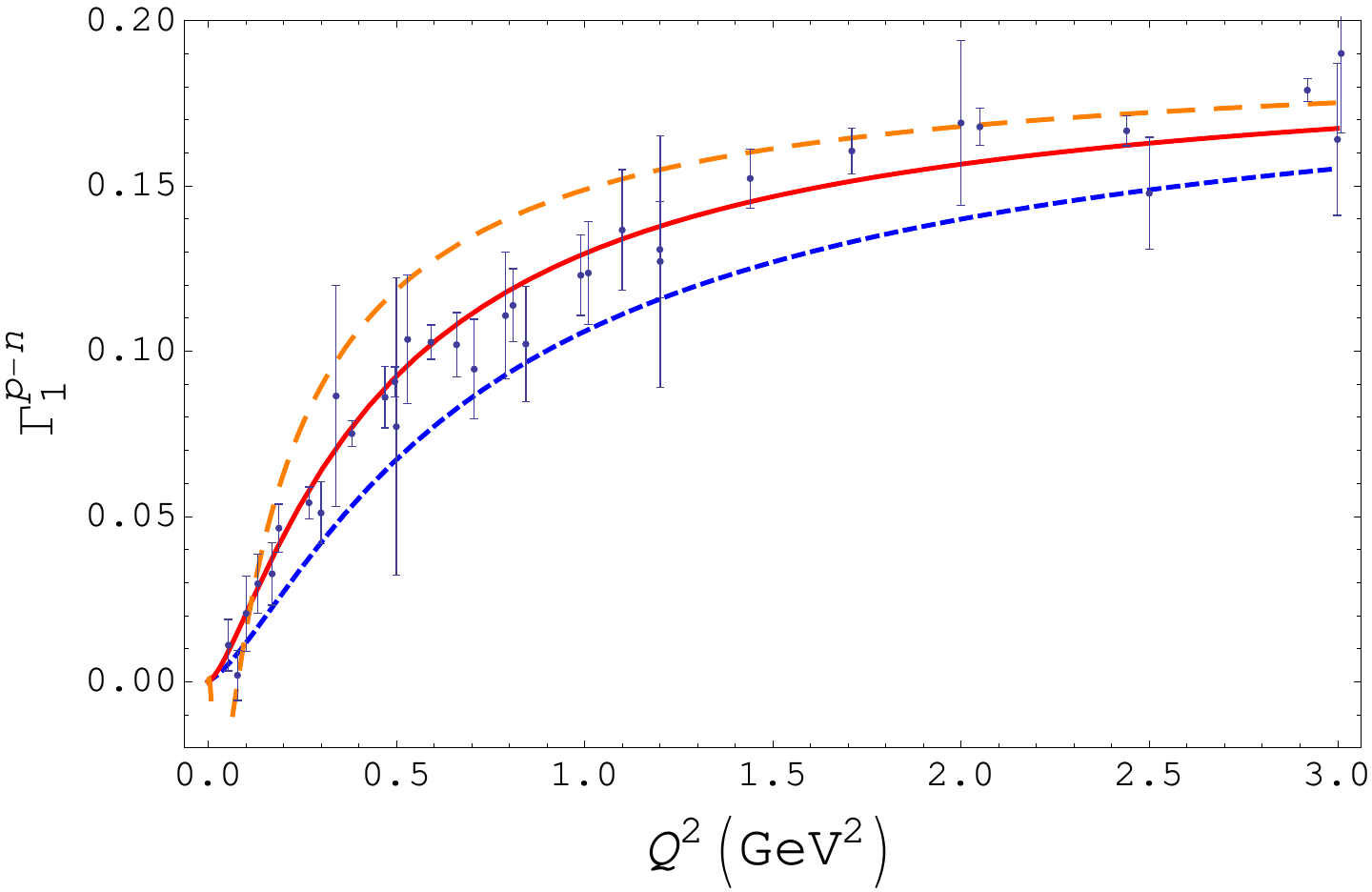}}
\caption{BSR at LO (dotted blue curve), NLO (solid red curve),
NNLO(dashed yellow curve)  compared with experimental data.}
\label{fig:1}
\end{figure}


It is well known that modified scaling variables provide the way of
partial HT resummation. This corresponds to famous Nachtmann scaling
variable and another one $\hat x = \tilde Q^2/2M \nu$, closely
related to MPT and used at low $x$ region. The advantage of the last
variable for BSR studies is that it is larger than $x_B$ so that
$g_(\hat x) < g_1(x_B)$ and one may hope to describe \cite{KT} the
decrease of $\Gamma^{p-n}_1(Q^2)$ at small $Q^2$   required by the
finiteness of photoabsorption cross-section. However, for larger
$x_B$ the spectral condition for $\hat x$ may be violated, so that
$\hat x \geq 1$. To avoid that, one may introduce another variable,
using the representation $x_B=Q^2/(Q^2+W^2)$ and performing the same
change $Q \to \tilde Q$:
\begin{equation}
\tilde x = x_B \frac{1+a}{1+a x_B},\,\, a=\frac{M^2_{gl}}{Q^2}.
\end{equation}
The first moment may be now presented as
\begin{equation}
\tilde \Gamma_1(Q^2) = \int_0^1 g_1(\tilde x) dx = (a+1) \int_0^1 \frac{g_1(z)dz}{(1+a(1-z))^2}.
\end{equation}
While asymptotically this result reproduces the standard partonic
expression, at smail $Q^2$ one get
\begin{equation}
\tilde \Gamma_1(Q^2) \to \frac{Q^2}{M^2_{gl}}\int_0^1 \frac{g_1(z)dz}{(1-z)^2},
\end{equation}
and the slope at the origin related to GDHSR is defined by the
second inverse moment. Note that the contribution of large x region
is enhanced. The available parameterizations of spin-dependent
parton distributions seem to provide a rather reasonable of data at
low $Q^2$ \cite{KT}.

\section{Hadronic vs vacuum matrix elements}
\subsection {Transverse Momentum Dependent Distributions and vacuum condensates}

The TMDs can be conveniently defined\cite{Teryaev:2004df}   in coordinate (impact parameter) space.
The representative case is Boer-Mulders function (in order to avoid consideration of gauge links/gluonic poles
one may consider\cite{Teryaev:2004df}  Collins fragmentation function having the same Lorentz structure) when transverse coordinate is selected by the chiral-odd Dirac structure
 \begin{equation}
\langle P | \bar \psi(0) \sigma^{\mu \nu} \psi(z)  |P \rangle = M (P^\mu z^ \nu-P^\nu z^ \mu) I(z\cdot P, z^2)
\label{x}
\end{equation}
 Here the $z^2$-dependence corresponds to $k_T$ dependence in momentum space and contains all twists. The definite twists may be extracted by the expansion in Taylor series
 \begin{equation}
I(z\cdot P, z^2)=\sum_{n=0}^\infty \frac{\partial^n}{n! \partial z^{2n}}I(z\cdot P, z^2)|_{z^2=0},
\end{equation}
which in the momentum space corresponds to the transverse moments of TMD. Note
that the lowest twist is 3, which is seen from the appearance of factor M in the r.h.s. of (\ref{x}).
In the momentum space this factor is shifted to the denominator, as the transverse moment is taken over $d k_T^2/M^2$ \cite{Teryaev:2004df}.

Note that the expansion in $z^2$ may be performed only after the subtraction of the singular terms
in $z^2$, which in the collinear factorization are absorbed to coefficient function. For TMDs
they constitute the power-like tail, after subtraction of which all the transverse moments became finite,
indicating the Gaussian distributions.

It is interesting that the appearance of such a tail was recently deduced \cite{Efremov:2013mpa}
from the causality arguments, being as it is well known also the origin of analytic properties discussed above.

The described situation is rather similar to  non-local vacuum condensates (see \cite{Bakulev:1991ps} and Ref. therein)
where one is dealing with vacuum matrix element
 \begin{equation}
\langle 0 | \bar \psi(0) \psi(z)  |0 \rangle = \langle 0 | \bar
\psi(0) \psi(0)  |0 \rangle F(z^2). \label{v}
\end{equation}
The Taylor expansion of F selects the local condensates of definite
dimension, corresponding to the moments of suitably chosen Fourier
transformed function being the complete analog of TMD. Note that
subtraction of singular  terms is performed here by subtraction of
the perturbative contribution corresponding to quark propagator.

Generally, the hadronic matrix elements differ from the vacuum ones
by the presence of essentially pseudo-Euclidian hadron momentum. At
the same time, the Euclidian transverse dynamics may be more
vacuum-like.

\subsection{D-term and cosmological constant}

Let us discuss one more interesting example of interplay between hadronic and vacuum matrix elements.
It corresponds to so-called Polyakov-Weiss D-term \cite{Goeke:2001tz} (appearing in analyticity based analysis as a subtraction constant \cite{Teryaev:2005uj})  whose moment is related to quadrupole gravitational formfactor
transverse coordinate is selected by the chiral-odd Dirac structure
 \begin{equation}
\langle P+q/2 | T^{\mu \nu} |P-q/2 \rangle = C(q^2)(g^{\mu
\nu}q^2-q^\mu q^\nu)+... \label{emt}
\end{equation}
where gravitoelectric and gravitomagnetic formfactors
\cite{Teryaev:1999su} are dropped. C has definite (positive) sign in
all the known cases  including hadrons \cite{Goeke:2001tz}(where
also general stability arguments are discussed), photons
\cite{Gabdrakhmanov:2012aa}, Q-balls \cite{Mai:2012yc}.

For vacuum matrix element one has the famous cosmological constant
 \begin{equation}
\langle 0| T^{\mu \nu} |0 \rangle = \Lambda g^{\mu \nu}
\label{c}
\end{equation}
One may relate this matrix element in 2-dimensional transverse space orthogonal to P and q, so that
effective 2-dimensional cosmological constant is
 \begin{equation}
\Lambda = C(q^2) q^2
\label{eff}
\end{equation}
 The positive C leads to negative cosmological constant in the scattering process and to positive one
 in the annihilation process. There seems to be some relation between annihilation and inflation!
 It may not be so surprising due to known similarity\footnote{I am indebted to A.A. Starobinsky for this comment.}
  between inflation and Schwinger pair production in the electric field.

It is of course very interesting whether real cosmological constant in our Universe may be understood
as emerging from annihilation at extra dimensions. Qualitatively this is similar to brane cosmology,
and one should stress that in the suggested scenario Big Bang is due to one-graviton annihilation.
The specification of extra-dimensional states providing the cosmological constant of mass dimension 4 remains to be investigated,

\section{Conclusions}

The analyticity property continues to play the major role in
developing of QCD approaches to low scale processes, being of most
experimental interest. It allows one to justify the representation
of infinite sums of higher twists contributions, providing, in
particular, the accurate description of Bjorken Sum Rule data at low
$Q^2$. Another method of higher twists partial resummation is
provided by modified scaling variables, sometimes allowing to
describe the real photon limit of DIS.

The infinite series of higher twists are required to transverse momentum dependent parton distributions, having deep similarity to non-local vacuum condensates.

The vacuum/hadron matrix elements similarity allows to describe one-graviton annihilation as effective 2-dimensional
cosmological constant, and this may be geberalized to describe in a similar way cosmological constant in our Universe
as emerging from one-graviton annihilation at extra dimension, which is the picture of a Big Bang in such a case.

I am grateful to Organizers for warm hospitality at HIgh Tatras and
to Participants for many exciting discussions. This work is
supported in parst by  RFBR grants 11-01-00182, 12-02-00613  and 13-02-01060.

\label{}




\nocite{*}
\bibliographystyle{elsarticle-num}
\bibliography{martin}


\end{document}